\documentclass[aps, pre,12pt, endfloats*, preprint]{revtex4-1}
\pdfoutput=1

\usepackage{amsmath}	
\usepackage{graphicx}   	
\usepackage{verbatim}   	
\usepackage{color}      	
\usepackage{subfigure}  	
\usepackage[utf8]{inputenc}
\usepackage{hyperref}

\newcommand{\cue}{\ensuremath{\boldsymbol{Q}}}
\newcommand{\ef}{\ensuremath{\boldsymbol{F}}}
\newcommand{\msd}{MSD}

\begin{document}
\title{A wave-mechanical model of incoherent neutron scattering\\ II. Role of the momentum transfer}
\author{Hans Frauenfelder$^1$}
\author{Robert D.\ Young$^2$}
\author{Paul W.\ Fenimore$^1$}
\affiliation{$^1$Los Alamos National Laboratory, Los Alamos, NM 87545 USA, $^2$Arizona State University, Tempe, AZ 85287 USA} 
\preprint{LA-UR-15-26544}

\date{18 August 2015 --- LA-UR-15-26544}

\begin{abstract}
We recently introduced a wave-mechanical model for quasi-elastic neutron scattering (QENS) in proteins.  We call the model ELM for ``Energy Landscape Model''.  We postulate that the spectrum of the scattered neutrons consists of lines of natural width shifted from the center by fluctuations. ELM is based on two facts:  Neutrons are wave packets; proteins have low-lying substates that form the free-energy landscape (FEL). Experiments suggest that the wave packets are a few hundred micrometers long. The interaction between the neutron and a proton in the protein takes place during the transit of the wave packet. The wave packet exerts the force $\ef(t) = d\cue(t)/dt$ on the protein moiety, a part of the protein surrounding the struck proton. $\cue(t)$ is the wave vector (momentum) transferred by the neutron wave packet to the proton during the transit. The ensuing energy is stored in the energy landscape and returned to the neutron as the wave packet exits. Kinetic energy thus is changed into potential energy and back. The interaction energy is proportional to 
$\cue$, not to $\cue^2$. To develop and check the ELM, we use published work on dehydrated proteins after reversing improper normalizations. In such proteins only vibrations are active and the effects caused by the neutron momentum can be studied undisturbed by external fluctuations.  ELM has predictive power. For example it quantitatively predicts the observed inelastic incoherent fraction $S(\cue, T)$ over a broad range of temperature and momentum $\cue$ with one coefficient if $S(0, T)$ is known.
\end{abstract}

\maketitle

\section{Background\label{background}}
The goal of our research is the exploration of protein dynamics. Life can only exist if its building blocks can 
move \cite{1}. A vast arsenal of techniques has been brought to bear on protein dynamics. Our own work 
started many years ago using the M\"ossbauer effect and flash photolysis. Recently we began to look at 
neutron scattering \cite{2}. In these experiments neutrons with wave vectors $\boldsymbol{q}$ hit the protein. The 
intensities $S(\cue, T)$ 
of the scattered neutrons are measured as a function of the 
temperature $T$ at different scattering angles, characterized by their wave vectors $\boldsymbol{q}^\prime$. The wave 
vectors $\boldsymbol{q}$ and $\boldsymbol{q}^\prime$ determine the transferred wave vector, $\cue = \boldsymbol{q} - \boldsymbol{q}^\prime$. We use the 
standard notation where the wave vector $\boldsymbol{q}$  is related to the momentum by $\boldsymbol{p} = \hbar \boldsymbol{q}$. For convenience 
we will call $\cue$ also ``momentum transfer'' or simply momentum. Neutron scattering is treated in excellent 
texts and reviews \cite{3,4,5,6,7}. The salient experimental observation is easy to describe: The scattered 
slow neutrons appear to show an elastic line and a broad quasi-elastic band. The currently accepted model 
explains the quasi-elastic band as being due to the Doppler effect caused by spatial motions of the target; 
we call it SMM for ``spatial motion model''. In the SMM the spectrum of the scattered neutrons is separated 
into two components, an elastic line and a quasi-elastic band. The quasi-elastic band is assumed to be 
composed of broad Lorentzians centered at the elastic line. In the simplest model, the quasielastic 
Lorentzians are attributed to motions of the target proton in a Gaussian well with mean-square displacement 
(\msd{}) $\langle x^2\rangle$. $S(\cue,T)$ is then given by the Debye-Waller (Lamb-M\"ossbauer) relation
\begin{equation}
S(\cue,T) = \exp\left\{ -\cue^2 \langle x^2(T)\rangle \right\}
\label{LMeqn}
\end{equation}
Here $\langle x^2(T)\rangle$ is the average \msd{} of the protons in the protein.  If Eq.\ \ref{LMeqn} does not fit the experimental data, ``non-Gaussianity'' and ``dynamic heterogeneity'' \cite{8,9,10,11} are invoked. 

A number of issues caused us to look critically at this model: We could not find a transparent description of the quasi-elastic neutron scattering (QENS) and of the physics behind the momentum transfer $\cue$. The separation of the spectrum into an elastic line and a quasielastic band made no sense to us. The existence of broad homogeneous Lorentzians centered at the elastic line is difficult to accept because the uncertainty relation would imply very short lifetimes. Eq. \ref{LMeqn} predetermines the explanation of the quasi-elastic scattering: QENS must be due to motions in real space to account for the large mean-square displacement $\langle x^2\rangle$. The focus on spatial motions may have been influenced by the early scattering theories that assumed the absence of low-lying substates\cite{12}. Non-Gaussianity and dynamical heterogeneity appear like first-aid patches to save the model.  These issues caused us to abandon the Gaussian approximation and its generalizations and search for a physics-based model with a testable role for $\cue$. The first step in this endeavor led to a model that does not involve Eq.\ \ref{LMeqn} and that is based on the free-energy landscape concept. We call it ``energy landscape model'' or ELM \cite{13,14}. It answers some questions, as we describe in the next section. The meaning of $\cue$, however, remained unclear. We propose to explain $\cue$ here. We therefore consider this paper part II of the PNAS publication
\cite{14}. 

\section{The Energy Landscape Model (ELM)}

The energy landscape model is based on two pillars: Neutrons are de Broglie wave packets\cite{15,16}; proteins have a large number of low-lying conformational substates (CS) that are organized in the free-energy landscape (FEL)\cite{17,18}. The wave nature of the neutron implies that the neutron-proton interaction time $\tau_n$ can be of the order of hundreds of ns. The existence of the CS implies that proteins can change their conformation and their energy during the passage of the wave packet. The two pillars permit the construction of the ELM as we have described previously\cite{13,14}. 
\begin{figure}
\centering
\includegraphics[scale=0.5]{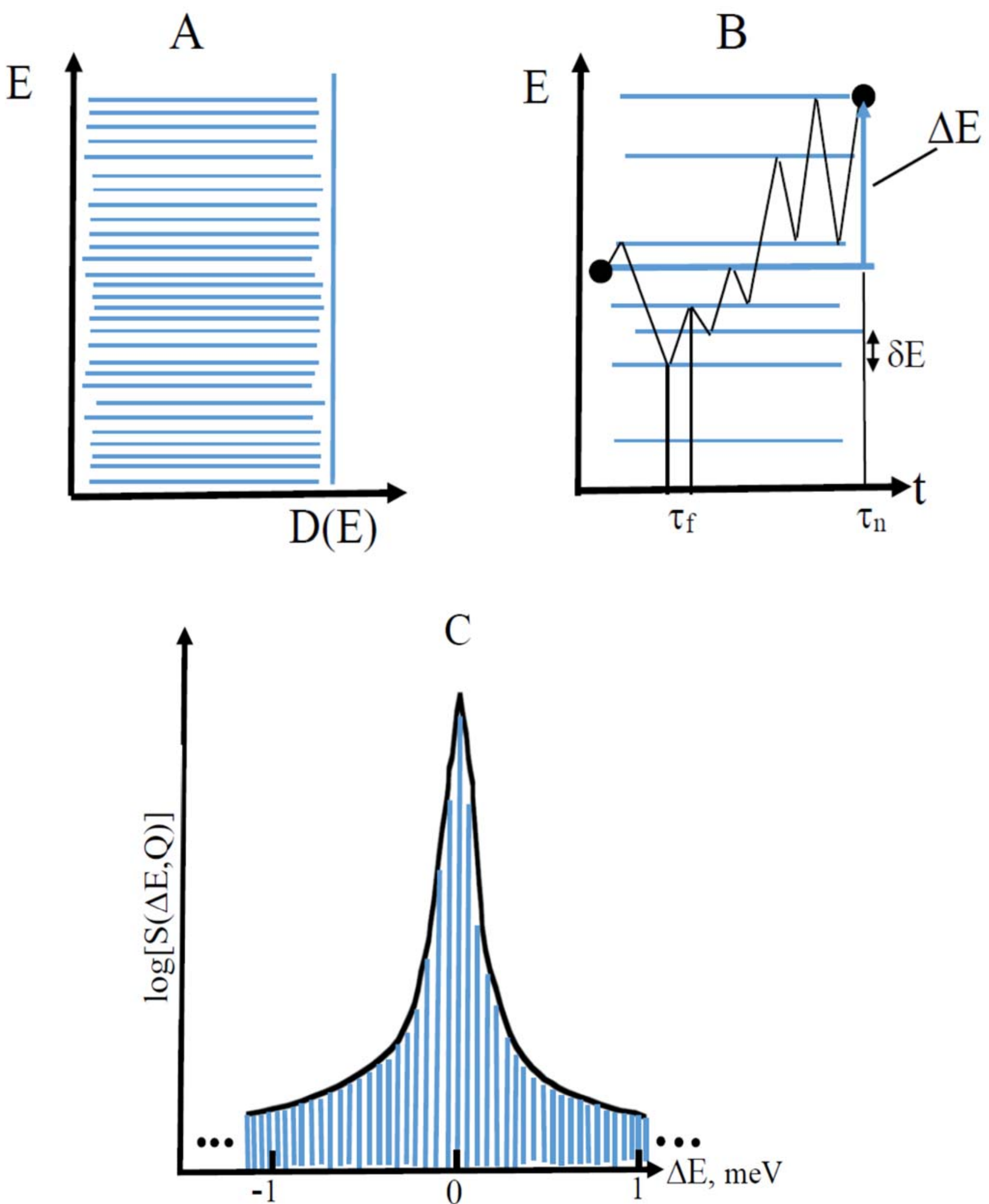}
\caption{\label{Fig1}(A) Hypothetical density of substates  D(E). (B) Random walk of a protein in the energy landscape, caused by thermal fluctuations. $\tau_f$ is a typical vibration frequency, $\delta E$ typical jump energy  between substates, and $\Delta E$ is the final substate energy shift. (C) $\Delta E$ has shifted a line from it initial position in the center line to its final position.}
\end{figure}
Fig.\ \ref{Fig1} describes the essential properties of the FEL. Fig.\ \ref{Fig1}(A) is a schematic of the density $D(E)$ of the substates in the FEL. $D(E)dE$ is the fraction of substates with energies between $E$ and $E + dE$. We assume that $D(E)$ is constant below about 50 meV. The population $\rho(E)$ of the substates depends of course on the temperature.  In Fig.\ \ref{Fig1}(B) a few CS have been selected. Consider a protein in a particular substate. In the absence of fluctuations it stays in this substate for a long time and neutron scattering yields a single line at $\Delta E_0 = 0$. The width of the line is given by the resolution function of the instrument. Proteins have a broad spectrum of thermal fluctuations. Below about 10K, the fluctuations are too slow to induce observable transitions among substates. As the temperature is increased, vibrations become faster and their density increases\cite{19,20}. The vibrations switch a protein from CS to CS, with average fluctuation time $\tau_f$. Assume that the system is observed for a time $\tau_n$. If $\tau_n$ is much shorter than $\tau_f$ no temporal information can be recovered. However, if $\tau_n$ is much longer than $\tau_f$ the energy shift during the time $\tau_n$ can be extracted as is suggested by Fig.\ \ref{Fig1}(B). The protein fluctuates approximately $\tau_n/\tau_f$ times during the observation time $\tau_n$, each time jumping from substate to substate and winding up in a substate shifted by the energy $\Delta E$. The intensity $S(0, T)$ of the peak decreases and a Lorentzian appears at $E + \Delta E$. The shifts $\Delta E$ can be positive or negative, and the result is an inhomogeneous spectrum, composed of shifted Lorentzians as sketched in Fig.\ \ref{Fig1}(C) The spectrum is dynamic; there is no difference between the central line and lines in the quasi-elastic band. No substate is permanently occupied; any protein can transiently be in any substate. There is no separation into an elastic line and a quasielastic band. All individual spectral lines have the same width, of the order of a few neV. The energy of the initial substate is irrelevant for $\Delta E$, only the change in energy during the time $\tau_n$ counts. Different proteins experience different $\Delta E$ and the ensemble produces a broad spectrum as in Fig. \ref{Fig1}(C). In neutron scattering the relevant time $\tau_n$ is determined by the optics in the neutron source\cite{15} and can be of the order of hundreds of ns.  In the M\"ossbauer effect, the lifetime $\tau_\mathrm{Mo} = $141 ns replaces $\tau_n$. In the M\"ossbauer effect the natural line width is extremely small and the decrease of the central intensity and the resulting line broadening can be observed already below 20 K \cite{21}. In neutron scattering, the observed line broadening is influenced strongly by the resolution function of the instrument and occurs at higher temperature than in the M\"ossbauer effect. 

Vibrations are not the only fluctuations that produce transitions among substates. In hydrated proteins the $\alpha$-fluctuations in the solvent and the $\beta_\mathrm{h}$-fluctuations in the hydration shell can cause protein motions \cite{22,23,24,25,26}. QENS studies of systems involving more than one type of fluctuations are harder to interpret and hide the simplicity of the ELM. We therefore use in the present paper only dehydrated proteins, where vibrations rule. 

\section{Forward Incoherent Neutron Scattering}
In principle, the fraction $S(0,T)$ can provide crucial information. In practice, however, it is difficult to measure it directly.
$S(0, T)$ must thus be determined indirectly. One approach is to average a number of measurements with small $\cue$ values as done by Hong et al.\cite{27}. This approach yields reasonable data, but for $\cue > 0$. We have used the data at different $\cue$ values in the supplementary information to the paper by Hong et al. to extrapolate $S(\cue, T)$ to $\cue = 0$. 
\begin{figure}
\centering
\includegraphics[scale=0.5]{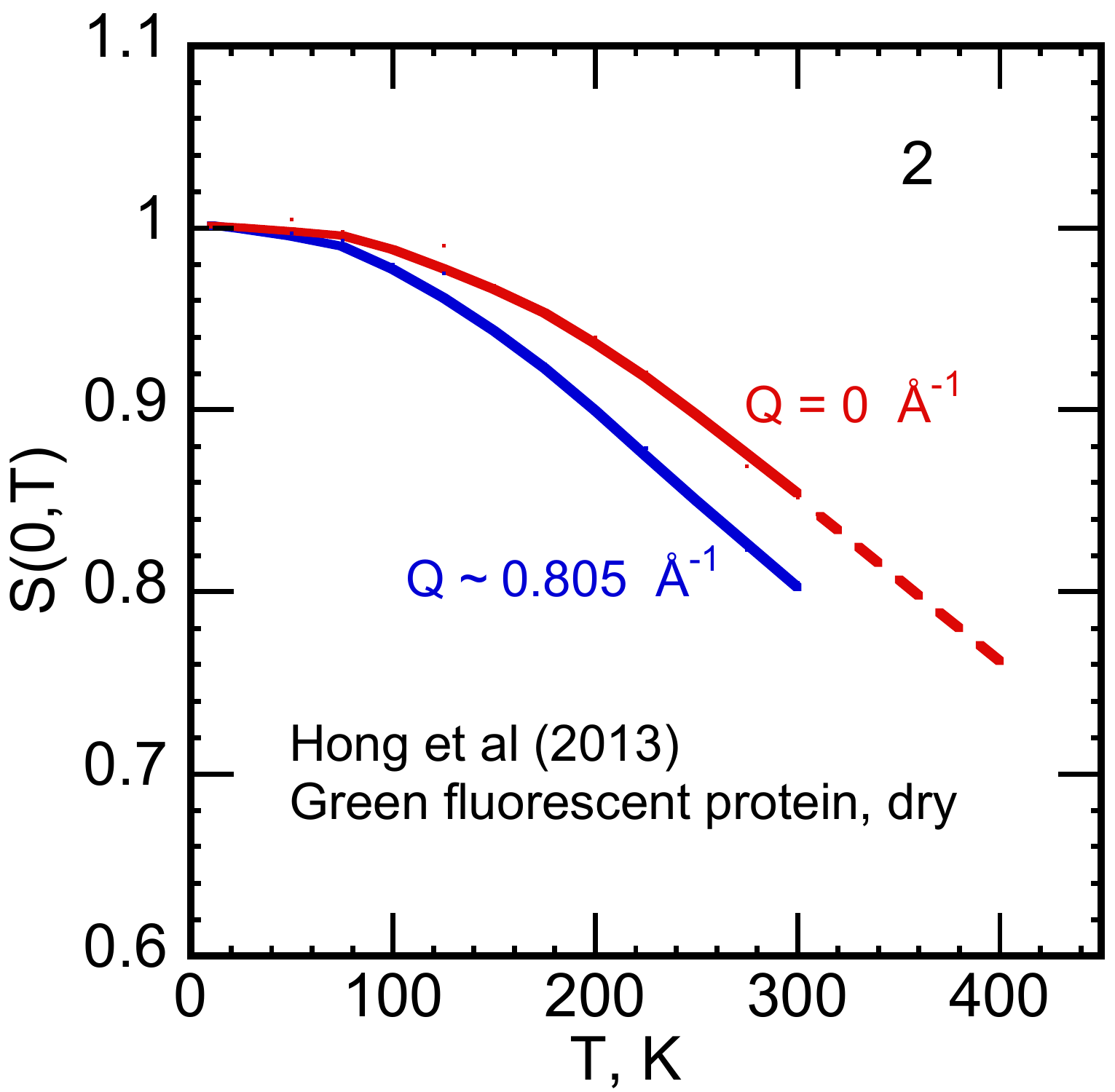}
\caption{\label{Fig2}The elastic fractions $S(\cue, T)$ for $\cue = 0 \mathrm{\AA}^{-1}$ and $\cue = 0.8 \mathrm{\AA}^{-1}$ for dry green fluorescent protein (GFP).  Data taken from Figure S1 of the Supplementary Material of Hong et al. \cite{27}. The curve for $\cue = 0$ was obtained by extrapolating the  data for three wave vector transfers to $\cue = 0$.  The dashed line results from an extrapolation between 250 and 300 K to 400 K.}
\end{figure}
The result, after normalizing $S(0, 0)$ to one, is given in Fig.\ \ref{Fig2}. Also shown in Fig.\ \ref{Fig2} is $S(0.8 \mathrm{\AA}^{-1}, T)$ to indicate that there is a measurable difference between $S(0.8\mathrm{\AA}^{-1}, T)$ and  $S(0, T)$. The reason why $S(\cue > 0, T)$ is smaller than $S(0, T)$ will become clear later. Fig.\ \ref{Fig2} also suggests that $S(0, T)$ decreases little below about 100 K. This effect is due to the broad resolution function in neutron scattering. The decrease in $S(0, T)$ becomes only observable when a sizable fraction of proteins have moved away from the center. However, comparison with M\"ossbauer data \cite{21}, where the energy resolution is much better, indicates that the decrease below 100 K is indeed small so that Fig.\ \ref{Fig2} is a good approximation. 

\section{Denormalization}
Normalization is the migraine headache of QENS. Improper normalization can warp the data. The correct normalization is $S(0, 0) = 1$. We did not find data approaching this ideal. Most data are ``normalized'' using one or sometimes two procedures: 
\begin{enumerate}
\item The assumption that all motions are absent at very low temperatures, for instance at $T_0 = 10\, \mathrm{K}$, leads to the normalization $S(\cue, T)/ S(\cue, T_0)$. This procedure is incorrect because quasi-elastic effects exist even at 10 K as we will show. Denormalization is done by multiplying the published data by $S(\cue, T_0)$ to recover $S(\cue, T)$. \label{first}
\item Assuming that the elastic effects can be removed without changing the quasi-elastic shape leads to division by the elastic fraction, $S(\cue, T)/S(0, T)$. This normalization changes the shape of the published $S(\cue, T)$. Denormalization is in principle  performed by multiplying the published $S(\cue,T)$ by the elastic fraction $S(0, T)$. If the $S(0, T)$ used in normalization is not given in the publications, we denormalize by using $S(0, T)$ from Fig.\ \ref{Fig2}. While this procedure may lead to some deviation from the true value, it corrects the data in the right direction. \label{second}
\end{enumerate}

\begin{figure}
\centering
\includegraphics[scale=0.5]{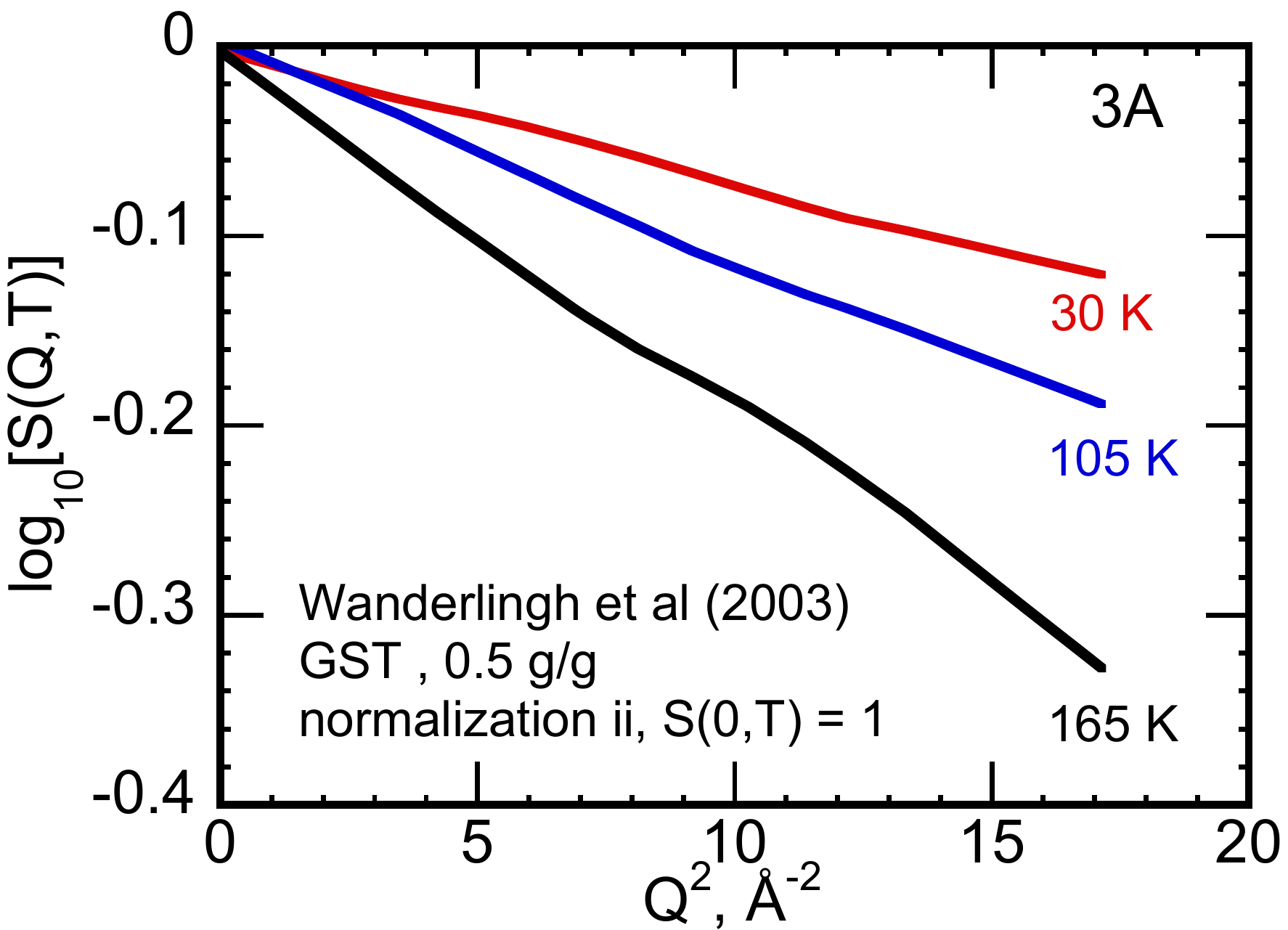}
\includegraphics[scale=0.5]{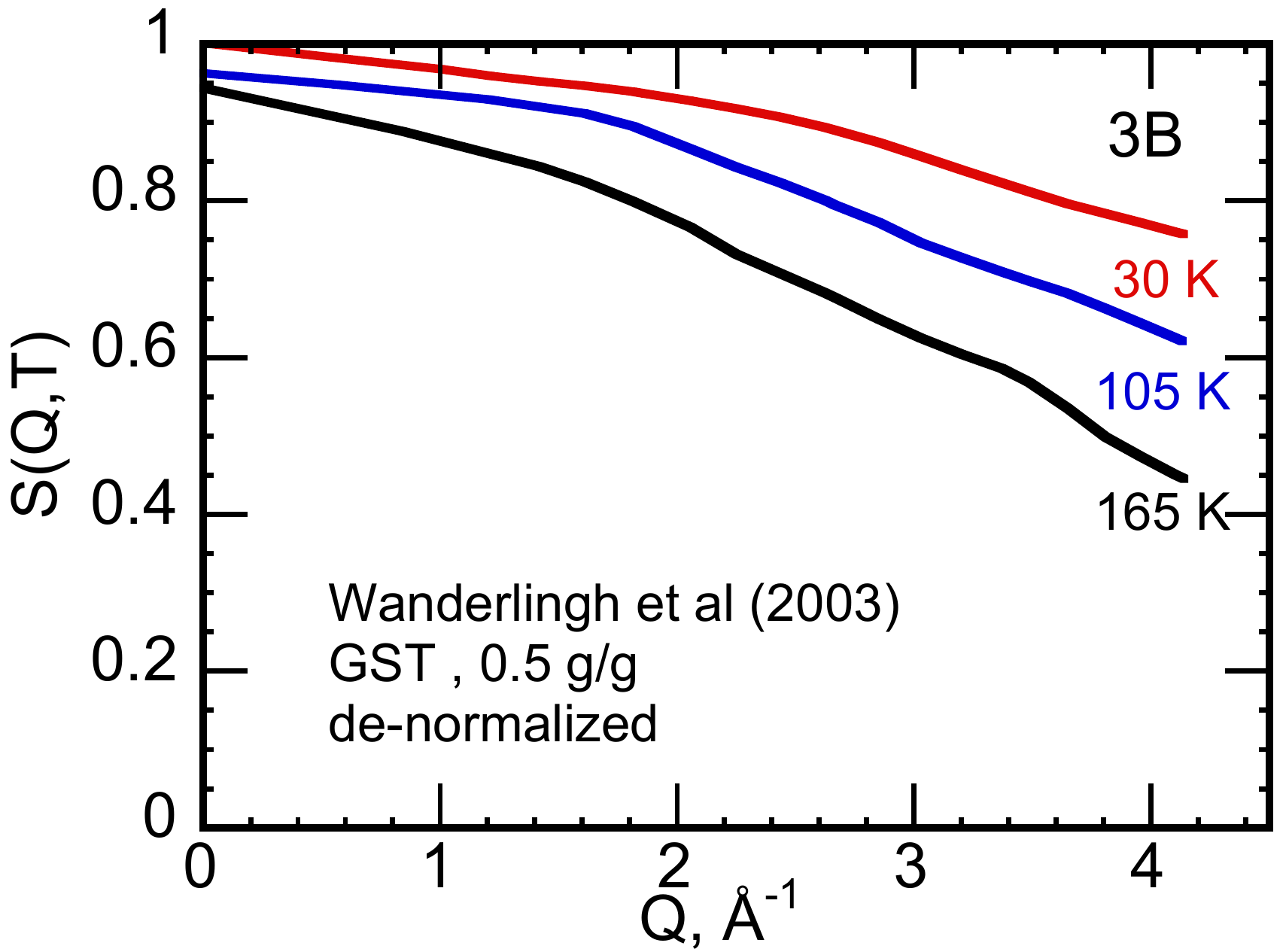}
\includegraphics[scale=0.5]{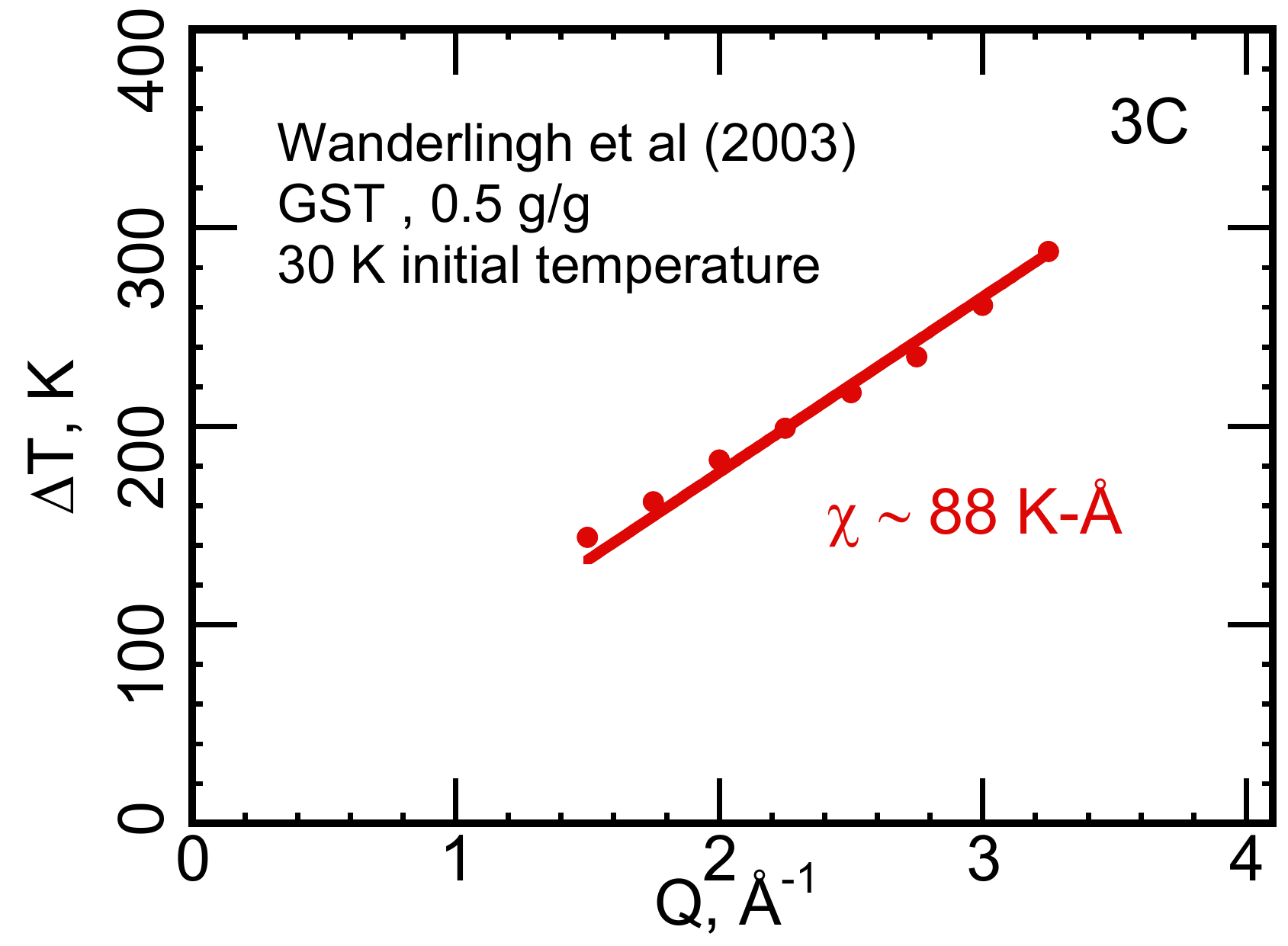}
\caption{\label{Fig3}(A) $\log (S(\cue,T))$ vs.\ $\cue^2$ for glutathione transferase (GST) at hydration of  0.5 and temperatures 30, 
105, and 165 K after smoothing of digitized data.  Adapted from Wanderlingh  et al. \cite{28} where the data is normalized 
to $S(0,T) = 1.0$.  (B)  $S(\cue, T)$ plotted versus \cue{} for the de-normalized data using Fig. \ref{Fig2}.  See text.  (C) The relation $\Delta T(\cue) 
= \chi Q$ from Fig. \ref{Fig2} and \ref{Fig3}A--B as described in text.  $\chi = 88\, \mathrm{K\, \AA}$.  Smoothing based on Lowess 
(Weighted) algorithm implemented in KaleidaGraph 4.5, Synergy Software}
\end{figure}

As an example of denormalization we select QENS data by Wanderlingh and collaborators \cite{28}. 
These authors measured incoherent quasielastic spectra of glutathione S-transferase (GST). Their spectrum, after digitizing and smoothing, is given in Fig.\ \ref{Fig3}(A). Denormalization \textit{i} is not needed, because the authors used a vanadium standard to get the correct $S(\cue, T)$ and did not divide by $S(\cue, T < 30 K)$. Denormalization \textit{ii} is required because the data were normalized to $S(\cue, T) = 1$. Because $S(0, T)$ is not given in their publication we use $S(0, T)$ from Fig.\ \ref{Fig2}. The denormalized $S(\cue, T)$ is plotted as function of $\cue$ in Fig.\ \ref{Fig3}(B). 

\section{The Momentum Transfer $\cue$ Excites the Proton Moiety}
The data of Wanderlingh et al. in Fig.\ \ref{Fig3}(B) expose the surprising fact that  the fraction $S(\cue, T)$ 
depends strongly on $\cue$ even at 30 K. For instance $S(4\mathrm{\AA}^{-1}, 30 K)$ is only 0.76. Motions at 30 K are 
only due to thermal vibrations. These decrease $S(0, T)$ by at most a few percent at 30 K as shown in 
Fig.\ \ref{Fig2}. Fig.\ \ref{Fig2} also shows that $S(0, T) = 0.76$ corresponds to a temperature 
$T^\star \approx 400\mathrm{K}$. Thus the momentum transfer $\cue = 4\mathrm{\AA}^{-1}$ increases the effective 
temperature by $\Delta T^\star(4\mathrm{\AA}^{-1} \approx 370 \mathrm{K})$. Calculation of $\Delta T^\star(\cue)$ for a 
number of momenta leads to Fig.\ \ref{Fig3}(C). This figure shows that $\Delta T^\star(\cue)$ is proportional 
to $\cue$ and suggests the relation
\begin{equation}
\Delta T^\star (\cue) = \chi Q
\label{proportional}
\end{equation}
Here $\chi$ is a coeficient with a value of $\approx 88\, \mathrm{K\,\AA}^{-1}$.

$T^\star$ cannot be  a real temperature. A heat spike would dissipate in ps. A simple model suggests  that the energy $\Delta E^\star \approx k_\mathrm{B}\Delta T^\star$ can be stored as potential energy in the FEL. Consider an ideal elastic spring with spring coefficient $k$. If a force $F$ compresses the spring by the distance $x$ the spring reversibly stores the energy $\frac{1}{2} k x^2$. In proteins, the basic features may be similar. The force could be due to the neutron wave packet and the moiety around the proton could act like an elastic spring. While a change in temperature would be shared by the entire sample, a structural change can be local. It could for instance correspond to a small change in position of a side chain close to the target proton. The excited moiety thus could be the mechanical analog of a thermal hotspot.

\begin{figure}
\centering
\includegraphics[scale=1.2]{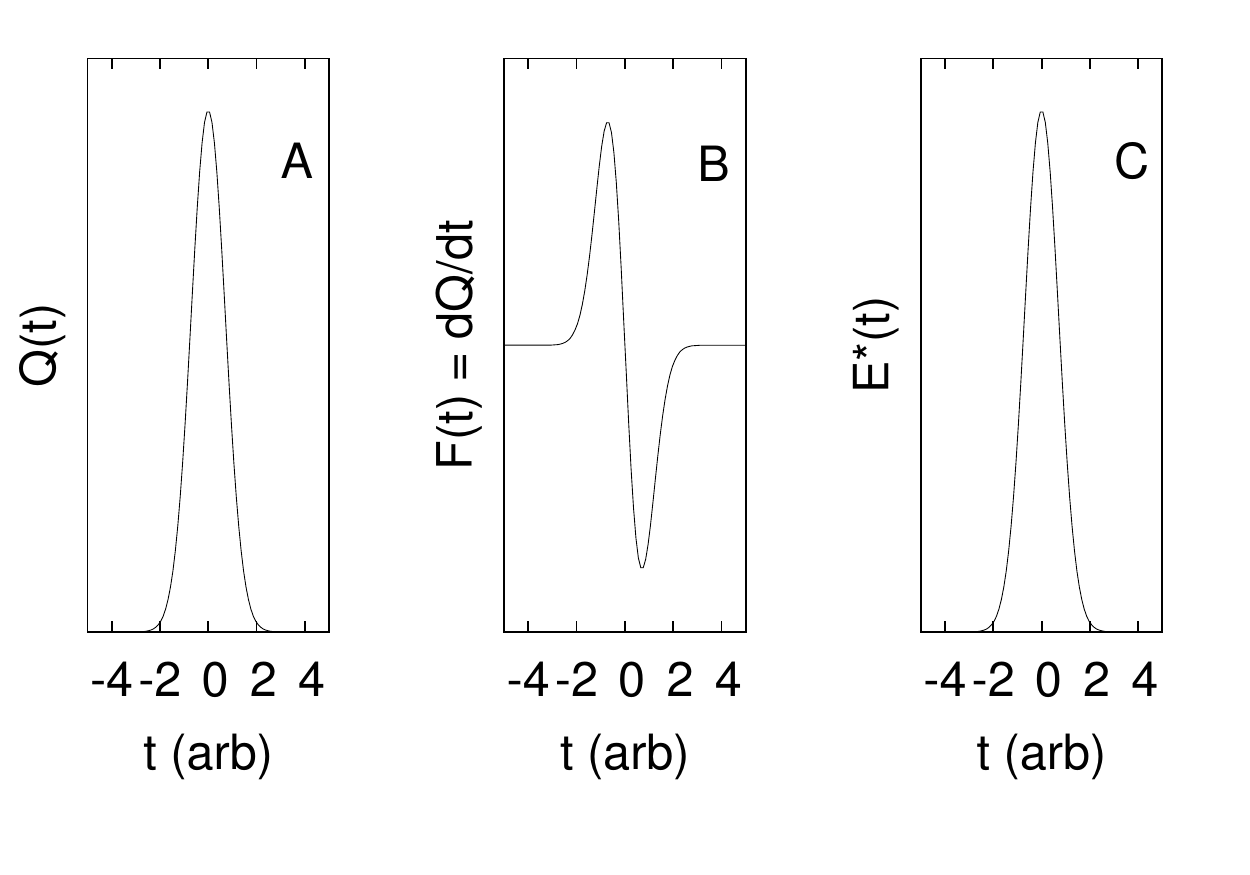}
\caption{\label{Fig4}(A) $\cue(t)$ is the intensity of the neutron wave packet at the proton at the time $t$. (B) The force $\ef(t)$ exerted by the wave packet on the proton at time $t$.  (C) The energy $\Delta E^\star(t)$ deposited in the proton moiety at the time $t$. }
\end{figure}
The suggested mechanism for the quasi-elastic neutron scattering is drawn in Fig.\ \ref{Fig4}. In Fig.\ \ref{Fig4}(A) the wave packet, approximated by a Gaussian, hits the proton with the momentum $\cue(t)$. The impact creates a force $\ef(t)$ given by,
\begin{equation}
\ef(t) = d\cue(t)/dt\, .
\label{force}
\end{equation} 
$\ef(t)$ is sketched in Fig.\ \ref{Fig4}(B). During the first phase of the transit, $\ef(t)$ is positive and the wave packet compresses the moiety by a distance $x$. This compression increases the potential energy $\Delta E^\star(t)$. During the second phase $\ef(t)$ is negative and the energy is returned to the wave packet. The protein exits with essentially the full energy, modulated by the quasielastic fluctuations that produce the broad band. The stored energy is sketched in Fig.\ \ref{Fig4}(C). The form of $\Delta E^\star(t)$ depends on the compressibility of the moiety. In the simplest case the process  leads to $\Delta E^\star \approx x^2$. In proteins the compressibility decreases markedly upon compression \cite{29} and the simplest relation may not apply. Fig.\ \ref{Fig4}(C) thus only gives a guess as to the form of the stored energy $\Delta E^\star(t)$ as a function of the transit time.

The energy input due to the momentum transfer could also be stored as kinetic energy in local vibrations.

Support for an elastic spring model comes from the fact that polymers can store energy \cite{30}. Shock wave experiments on polymer solids also involve energy storage \cite{31, 32, 33, 34, 35}. In an experiment by Dlott and coworkers a 50 $\mu$m thick aluminum plate hits a solid polymer target with a velocity of 1 km/s. The shock compresses the polymer solid. The deposited energy excites the FEL of the polymer, leading to an increase in the local temperature by about 100 K \cite{36}.

The observation that $\Delta T(\cue)$ in neutron scattering is of the same order as seen in the shock wave  experiments permits an estimate of the size of the target in the neutron scattering. In the shockwave experiment the momentum density is about $P_s = 10^2\, \mathrm{kg}\, \mathrm{m}^{-1}\, \mathrm{s}^{-1}$. In the QENS experiment the relevant momentum $P_n = m_n v_n$   is about $10^{-25}\, \mathrm{kg}\, \mathrm{m}\, \mathrm{s}^{-1}$. We interpret the ratio $P_n/P_s \approx 10^{-27}\, \mathrm{m}^2$ as the cross section for the neutron-proton scattering. It is remarkably close to the experimental value \cite{37} of $2.4 \times 10^{-27}\, \mathrm{m}^2$ and supports the assumption that the neutron interacts with a single proton. 

The comparison between the shock wave and the QENS processes cannot be taken too far. In the shock wave experiment, the aluminum plate is destroyed and the energy is ultimately dissipated as heat. In neutron scattering the wave packet interacts with a single proton that then puts pressure on its immediate surrounding. The pressure in turn changes the conformation of the proton moiety  \cite{38} thereby storing the energy reversibly in the FEL. The interaction is local and the FEL is only changed locally. The energy is not lost but returned to the neutron.

\section{The Effect of Vibrations}
Conformational motions can be caused by thermal vibrations, $\beta_h$-fluctuations in the hydration shell, or $\alpha$-fluctuations in the environment. If they are simultaneously present it is difficult to unravel their effects. We consequently selected a dehydrated protein where thermal vibrations dominate. We use QENS data of Nakagawa, Kamikubo, and Kataoka on dehydrated Staphylococcal nuclease (SNase) \cite{39}. 
\begin{figure}
\centering
\subfigure{\includegraphics[scale=0.4]{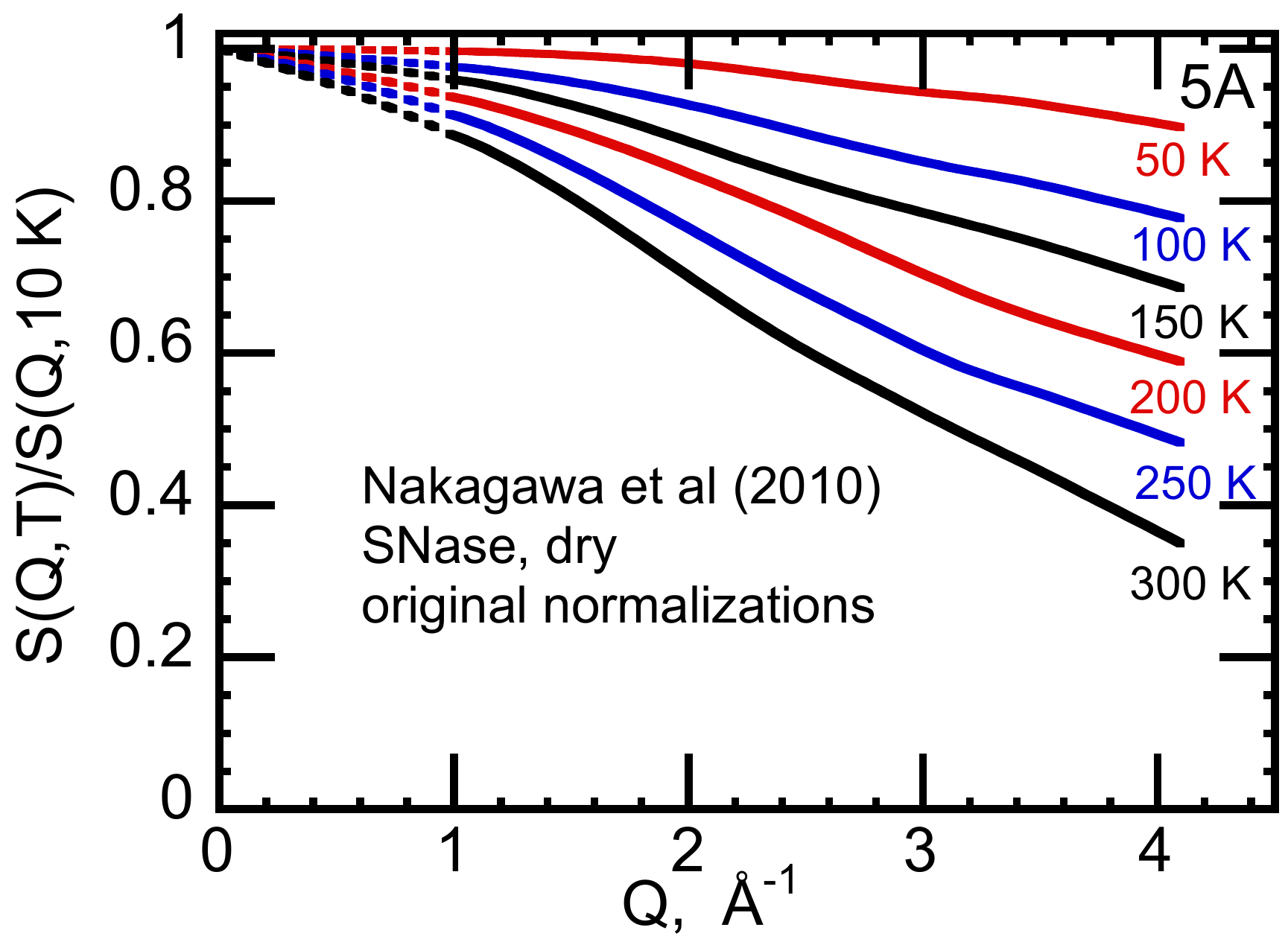}}
\subfigure{\includegraphics[scale=0.4]{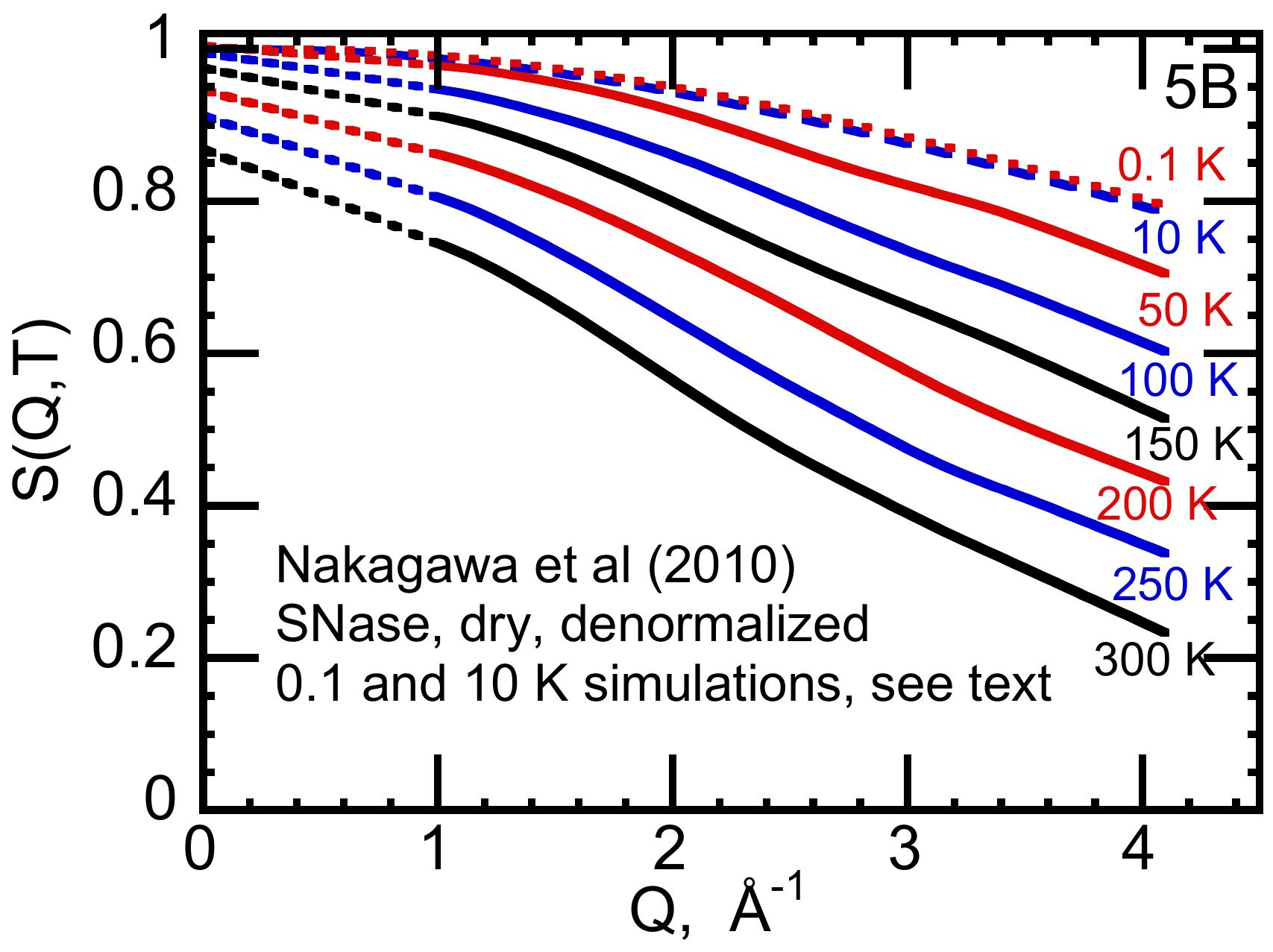}}
\subfigure{\includegraphics[scale=0.4]{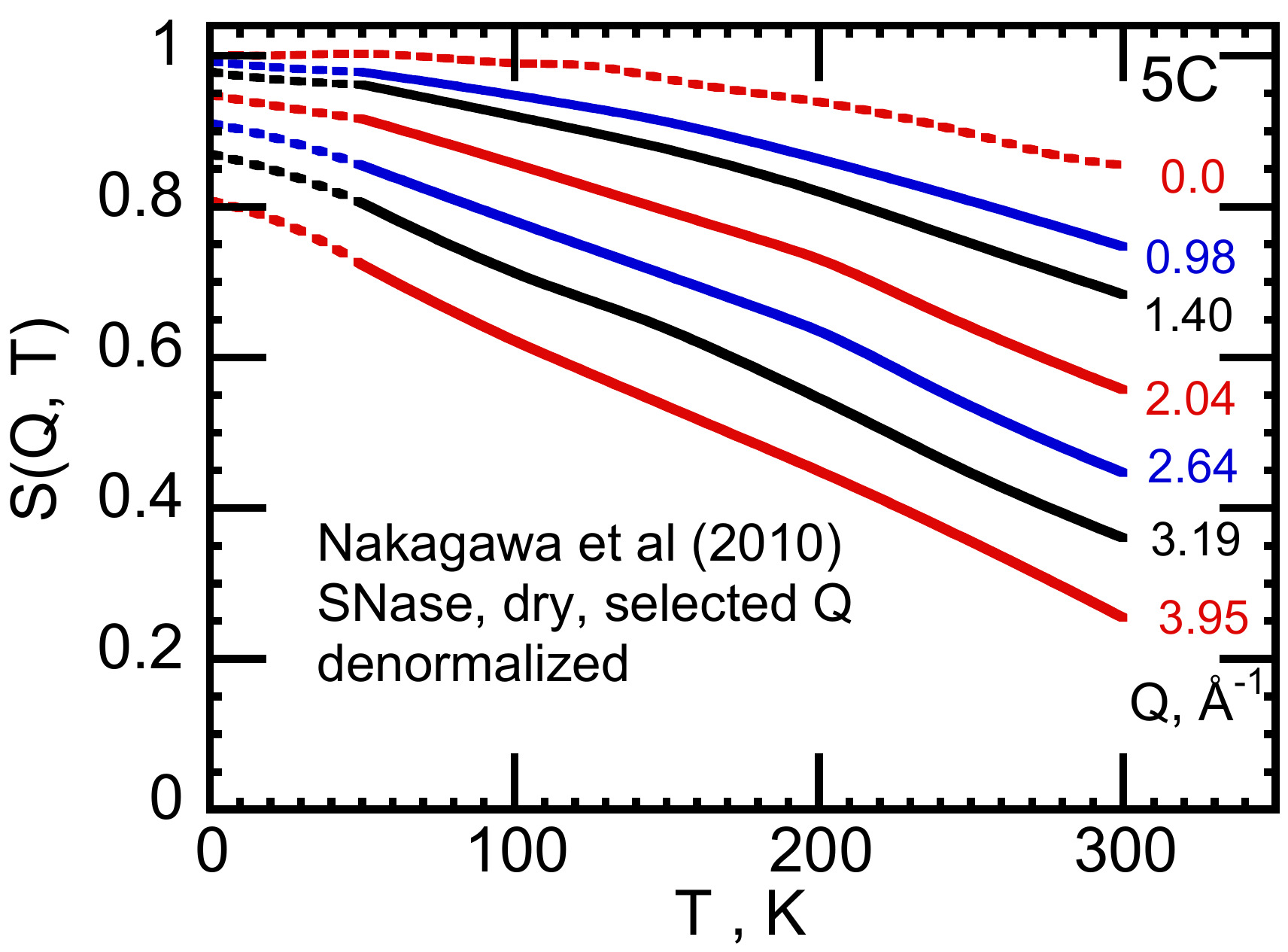}}
\subfigure{\includegraphics[scale=0.4]{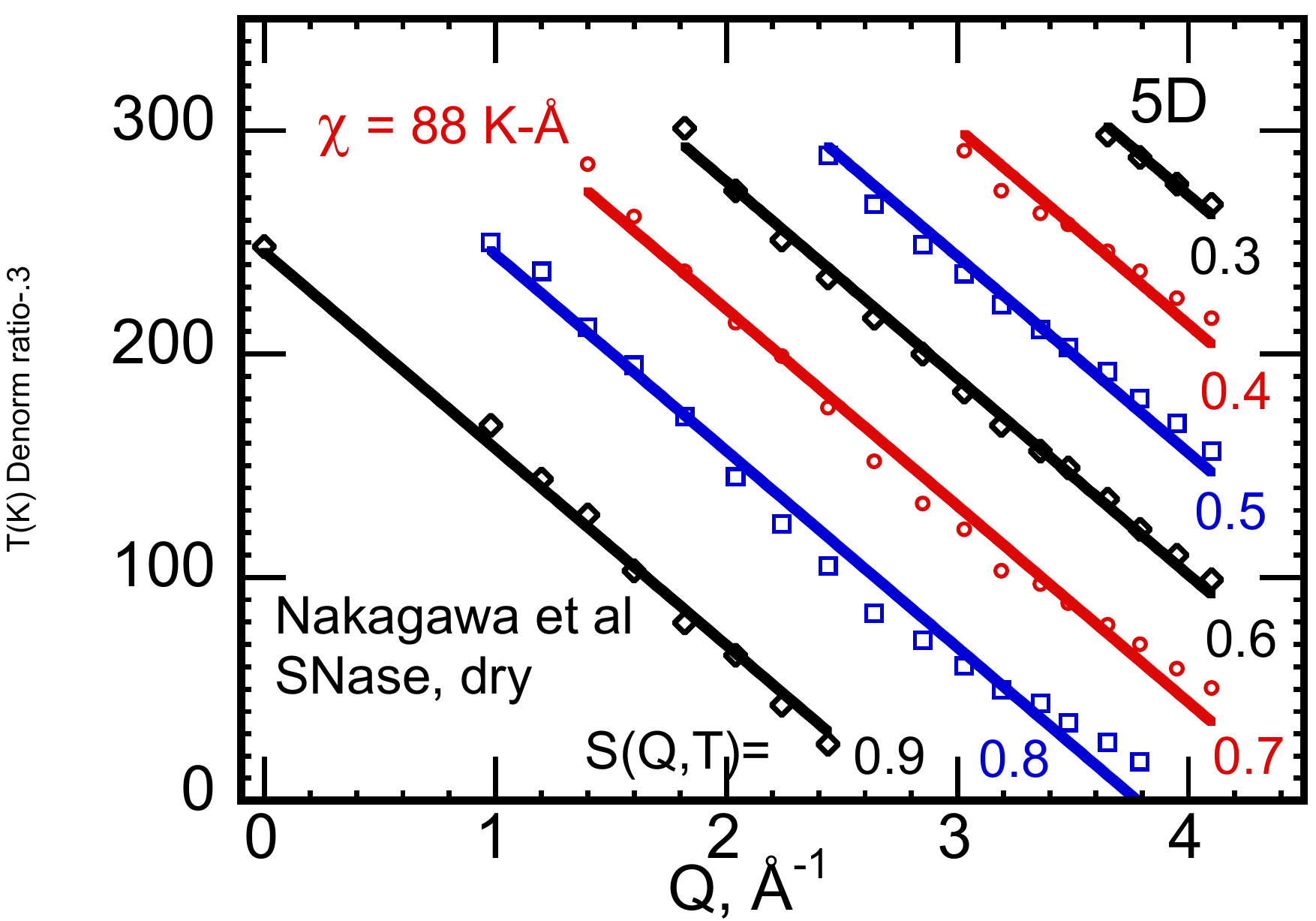}}
\caption{\label{Fig5}(A) Normalized $S(\cue,T)$ vs $\cue$ for dry staphylococcal nuclease (SNase).  Dashed lines show linear extrapolation from smallest measured $\cue$ to $\cue = 0$.  Adapted from Nakagawa et al \cite{39}.  (B) $S(\cue, T)$ vs $\cue$ for dry SNase after de-normalizations \textit{i} and \textit{ii}.  Dashed lines are linear extrapolations from smallest measured $\cue$ to $\cue = 0$.  Simulations of $S(\cue,T)$ at $T =$ 0.1 and 10 K (all dashed/dotted ) as explained in text.  (C) $S(\cue, T)$ vs $T$ for dry SNase including the curve for $\cue = 0$ (dotted).  Smoothing in (A), (B), and (C) is based on the Stineman (Smooth) and Lowess (Weighted) algorithms implemented in KaleidaGraph 4.5, Synergy Software, 2013.  (D)  Fig.\ \ref{Fig5}C shows that the intensities $S(\cue, T)$ at different values of $\cue$ have the same overall temperature dependence, but are shifted to lower temperatures with increasing $\cue$. To quantify the shifts, Fig.\ \ref{Fig5}D presents cross sections through the data in Fig.\ \ref{Fig5}C at selected values of $S(\cue, T)$ from 0.3 to 0.9. The cross sections are parallel with  $\chi\approx 88\, \mathrm{K \, \AA}$ as for glutathione S-transferase.}
\end{figure}
Fig.\ \ref{Fig5}(A) presents some of their data after digitizing and smoothing. The data imply that both normalizations \textit{i} and \textit{ii} have been used. Denormalization \textit{i} is straightforward; the data are multiplied by $S(0, T)$  from Fig.\ \ref{Fig2}.  Denormalization \textit{ii} is more involved because the measured $S(\cue, 10 \mathrm{K})$ has not been published. However, the data in Fig.\ \ref{Fig3}(B), together with Fig.\ \ref{Fig2}, permit one to reconstruct $S(\cue, 10\mathrm{K})$. We assume that Eq.\ \ref{scaling} holds even at the lowest temperature. Fig.\ \ref{Fig3}(C) predicts that the temperature increase for instance at $\cue = 4 \mathrm{\AA}^{-1}$ is about 352 K and $T^\star = 362 \mathrm{K}$. Fig.\ \ref{Fig2} then gives $S(4 \mathrm{\AA}^{-1}, 10 K) \approx 0.79$.  Thus values for $S(\cue,T)$ can be found even for $T = 10 \mathrm{K}$, and denormalization \textit{ii} can be performed. Fig.\ \ref{Fig5}(B) gives the denormalized $S(\cue, T)$.

Information on protein dynamics is contained in the temperature dependence of $S(\cue, T)$. This dependence is difficult to see in Fig.\ \ref{Fig5}(B). We therefore redraw $S(\cue, T)$ as function of temperature in Fig.\ \ref{Fig5}(C). Also shown in Fig.\ 5(C) is $S(0, T)$, taken from Fig.\ \ref{Fig2}.  The result is surprisingly simple: The $S(\cue, T)$ curves have essentially the same shape as the elastic $S(0, T)$, but are shifted to lower temperatures. This shift is at first counterintuitive. Why should the energy input due to the neutron lead to lower temperatures? The answer is in the difference between $T$ and $T^\star$. $T$ is the ambient temperature of the entire sample.  $T^\star$ gives the energy of the excited state in the FEL produced by the pressure of the neutron wave packet. $T^\star$ is higher than $T$ by $\Delta T^\star(\cue)$ given in Eq.\ \ref{force}. Thus the protein moiety measured at the low temperature $T$ is actually at the temperature (energy) $T^\star$.

To quantify the shifts between $T$ and $T^\star$ we plot in Fig.\ \ref{Fig5}(D) the temperatures of $S(\cue, T)$ at fixed values of $S(\cue, T)$. The curves are parallel and show that for a fixed S(\cue, T) the shift is proportional to $\cue$ and given by Eq.\ \ref{proportional} with $\chi = 88 \mathrm{K\, \AA}$. The neutron increases the temperature (energy) of the proton moiety by $\Delta T^\star(\cue) = \chi Q$, or
\begin{equation}
S(\cue,T^\star) = S(0, T+\chi Q)\, .
\label{scaling}
\end{equation}
The incoherent elastic fraction $S(0, T)$ and the coefficient  $\chi$  together predict $S(\cue, T)$ for all values of $\cue$.  $S(0, T)$ with dehydrated proteins is given by the spectrum of thermal fluctuations. The only fit parameter is the coefficient $\chi$ in Eq.\ \ref{proportional}.

\section{Conclusions}
Incoherent quasi-elastic neutron scattering (QENS) has traditionally been described as follows: Above a certain temperature, the sharp elastic line decreases in intensity and a broad ``quasi-elastic'' band, centered at the elastic line, appears.
In the conventional approach to neutron scattering, used for more than fifty years, the two phenomena are ascribed to spatial motions, for instance spatial diffusion \cite{3, 4}. No clear interpretation of the effect of the momentum transfer $\cue$ is given.

In ELM, the energy landscape model, the central line and the broad band are part of one spectrum, consisting of lines with essentially the natural width, and shifted by fluctuations. The neutron is a wave packet, a few hundred $\mu$m long. During the transit, the wave packet exerts the force $\ef(t) = d\cue(t)/dt$ on the proton moiety. The ensuing energy $\Delta E^\star(t)$ is temporarily stored in the free energy landscape. Thus the kinetic energy of the neutron is transformed into potential energy, not heat.  Some or nearly all of the energy is returned to the outgoing wave packet. The scattered neutron has essentially the full energy. During the passage vibrations induce transitions among the substates of the FEL, thereby producing the quasielastic band. To test the model, we have used published data after denormalization. Despite the fact that ELM explains the data, the currently accepted model must also contribute. Fluctuations in the energy landscape are connected to conformational motions in real space. A future theory must connect the two views.

The ELM solves or avoids the problems that we listed in Section \ref{background}. The ELM provides a physics-based explanation of QENS. The central line and the broad band are components of the same spectrum. The components are narrow and do not cause problems with the uncertainty relation.  Non-Gaussianity and dynamic heterogeneity are not needed.  The role of $\cue$ is clear; it creates the force that the passing neutron wave packet exerts on the proton and that leads to the excitation of the protein moiety. The effect of the wave packet is proportional to $\cue$, not to $\cue^2$. Spatial diffusion, which is difficult to explain at very low temperatures, does not enter. The observed spectra with non-zero $\cue$ are quantitatively predicted once $S(0, T)$ and the coefficient $\chi$ are known.  
The simplicity of the model is bought, however, with new problems. Can the neutron wave packet slow down and then speed up in its passage? Can the force due to the neutron wave packet reversibly excite the proton moiety without dissipation? What is the nature of the coefficient $\chi$? 

Finally a remark about the theoretical underpinning of the interaction of the wave packet with the proton. How can the long-lasting interaction be described by the instantaneous first Born-approximation? Baym and Pethick have treated this problem by assuming an instantaneous contact interaction and replacing the customary plane waves with wave packets \cite{40}.

\section{acknowledgments}
Part of the work was performed at Los Alamos under U.\ S.\ Department of Energy contract DE-AC52-06NA25396.

Frequent interactions with many colleagues have been important for the formulation of the present model. We thank in particular G.\ Baym, D.\ Dlott, J.\ N.\ Ginocchio, G.\ R.\ Kneller, S.\ Magazu, E.\ Mamontov, B.\ H.\ McMahon, F.\ Migliardo, A.\ Paciaroni, C.\ Pethick, D.\ Pines, and P.\ G.\ Wolynes.

\printfigures

\end{document}